\documentclass[prb,twocolumn,showpacs,preprintnumbers,superscriptaddress]{revtex4-1}
\usepackage[usenames,dvipsnames]{color}
\usepackage{amssymb} 
\usepackage{times}
\usepackage{graphicx}
\usepackage{graphicx}
\usepackage{dcolumn}
\usepackage{datetime}
\usepackage{soul}
\usepackage{subfigure}
\usepackage[bookmarks, colorlinks=true, plainpages = false,citecolor = red, urlcolor = black, 
filecolor = blue]{hyperref}
\usepackage{multirow} 
\usepackage[english]{babel}
\usepackage{graphicx}
\usepackage{amsmath}
 
\usepackage[english]{babel}
\usepackage{graphicx}
\usepackage{amsmath}
 
\begin{document}
\title{How to Make a Pristine Tellurium Atomic Helix}
 
\author{George Kirczenow}  

\affiliation{Department of Physics, Simon Fraser
University, Burnaby, British Columbia, Canada V5A 1S6}

\date{\today}

\begin{abstract}\noindent
Tellurium atomic helices are currently attracting increasing attention. However, to date individual tellurium atomic helices have only been grown encapsulated in host materials. A bare 
free-standing tellurium atomic helix has yet to be realized experimentally despite the fundamental interest of these systems. Here DFT-based simulations are presented that show that a pristine tellurium atomic helix can be drawn from a tellurium crystal with the help of a gold STM tip,  paving the way to a better understanding of tellurium atomic helices.
\end{abstract}

 \maketitle

 \section{Introduction}
\label{Intro}

Tellurium crystals consist of helical chains of atoms. Individual Te atomic helices encapsulated in various host materials have been synthesized. \cite{Bogomolov83MO,Bogomolov85MOZEO,Poborchii95MO,Inoue2005ZEO,KobayashiCNT,MedeirosCNT,IkemotoCNT,QinCNTBNNT,PoborchiiZEOL}  Studies of pristine (not encapsulated) single Te atomic helices would also be of interest\cite{ChurchillExfol} and are necessary for a complete understanding of Te helices. However, to date such bare tellurium atomic helices have not been realized experimentally. The purpose of this note is to point out how this may be accomplished.

Scanning tunneling microscope (STM) tips have been used  to image and rearrange atoms on surfaces,\cite{Eigl} to contact individual molecules,\cite{Bumm} to form atomic chains bridging electrical contacts,\cite{Ohn} and to perform electrical and thermal transport measurements on such systems.\cite{review} The computer simulations presented here show that it should be possible to make free-standing pristine Te atomic helices by  drawing them out from the surface of a Te crystal with the help of a gold STM tip. Te atomic helices made in this way would bridge a pair of electrical contacts (the gold STM tip and p-type semiconductor Te substrate) and thus be accessible  to direct transport measurements.

 \section{Method}
\label{M}

The computer simulations presented here were carried out  with the Amsterdam Modeling Suite (AMS 2024) using the  Perdew-Burke-Ernzerhof generalized gradient approximation density functional, the ZORA/TZ2P basis and including relativistic effects at the level of spin-orbit coupling. The atomic geometries were optimized requiring the energy gradients to be 10$^{-4}$ Hartree/Bohr or less. 

 \section{Results}
\label{R}

\begin{figure}[t]
\centering
\includegraphics[width=0.6\linewidth]{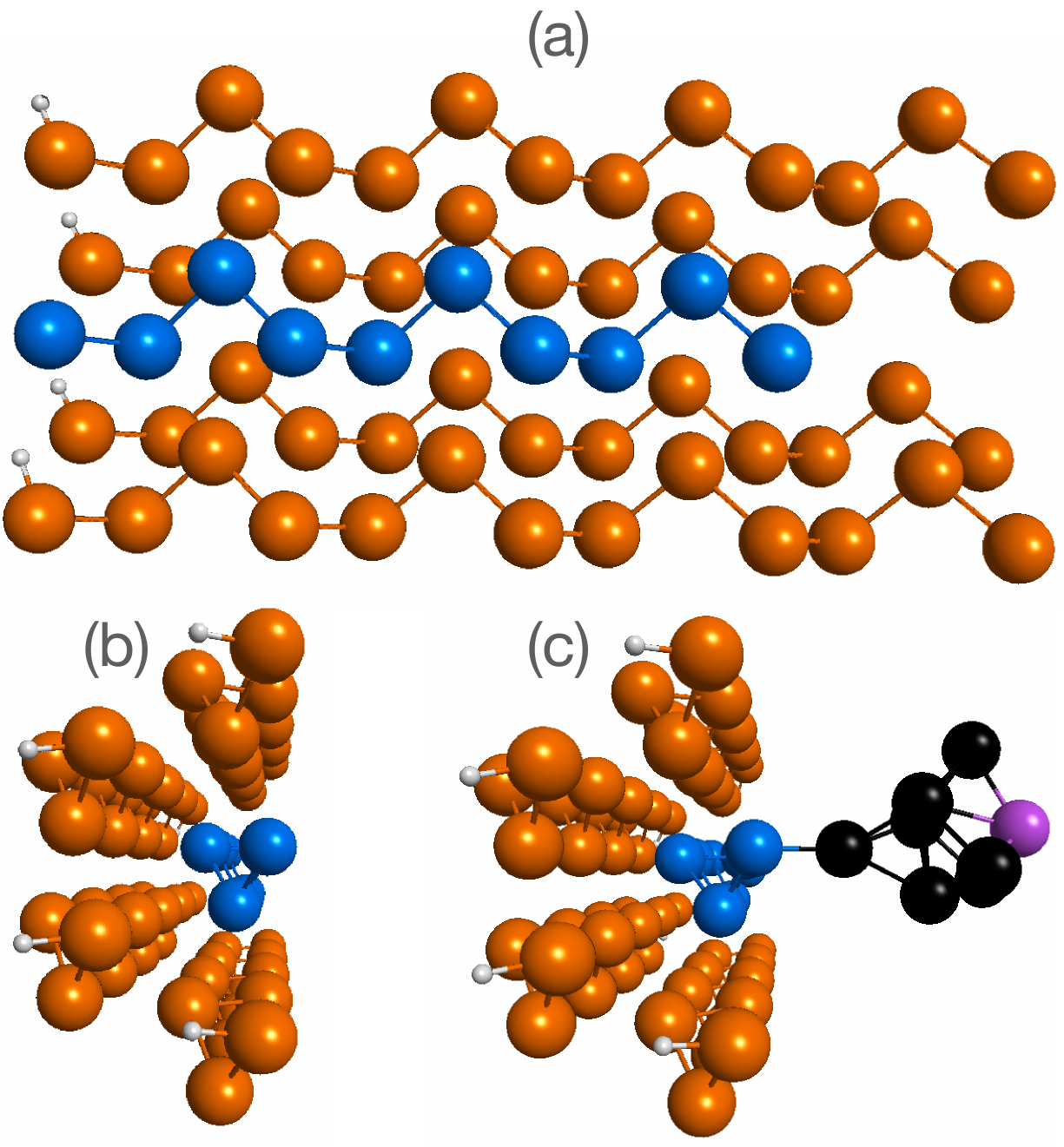}
\caption{(Color online).
(a) Top view and (b) side view of the Te atomic chain (blue) to be drawn out from the Te crystal surface. The Te atomic chains of the crystal that are adjacent to the blue chain  are also shown and are colored orange. They are terminated at both ends with H atoms (gray) to passivate dangling bonds. (c) Optimized geometry: 8 atom gold STM tip bonds to the Te atom at the right end of the (blue) Te chain in (a).  Gold atoms shown in black and Te atoms shown in blue are not constrained  during optimization while the mauve gold atom and orange Te atoms are frozen.  Images prepared using Macmolplt software. \cite{MacMolPlt}
}
\label{start} 
\end{figure}

A starting geometry is shown in Fig.\ref{start}
where the helical Te chain to be drawn out of the Te crystal surface is shown in blue and the adjacent Te helices are shown in orange. The proposed  procedure is as follows: The STM locates an end defect of the Te crystal surface where a Te helix ends, such as at the right hand end of the blue helical chain in Fig.\ref{start} (a). Then the gold STM tip atom shown in black in Fig.\ref{start} (c) is brought close to the end atom of the (blue) Te atomic chain and bonds to it. The resulting optimized geometry is shown in Fig.\ref{start} (c) where  the gold atoms shown in black and the blue Te atoms were not constrained  during optimization while the mauve gold atom and orange Te atoms were frozen. The orange Te atoms were frozen in the simulations in order to take into account the fact that they are held in place by their interactions with (not shown) adjacent helical Te atomic chains that extend over macroscopic distances. Of the gold atoms only the position of the one shown in mauve is frozen so as to control the overall height of the STM tip while allowing the positions of the other (black) tip atoms to relax fully.   The orange Te atomic chains in Fig.\ref{start} are terminated with H atoms so as to avoid the presence of reactive  dangling bonds that would not exist in this vicinity if the orange Te chains were part of a macroscopic crystal surface and  thus would extend far beyond the region shown in the figure.

\begin{figure}[b]
\centering
\includegraphics[width=0.6\linewidth]{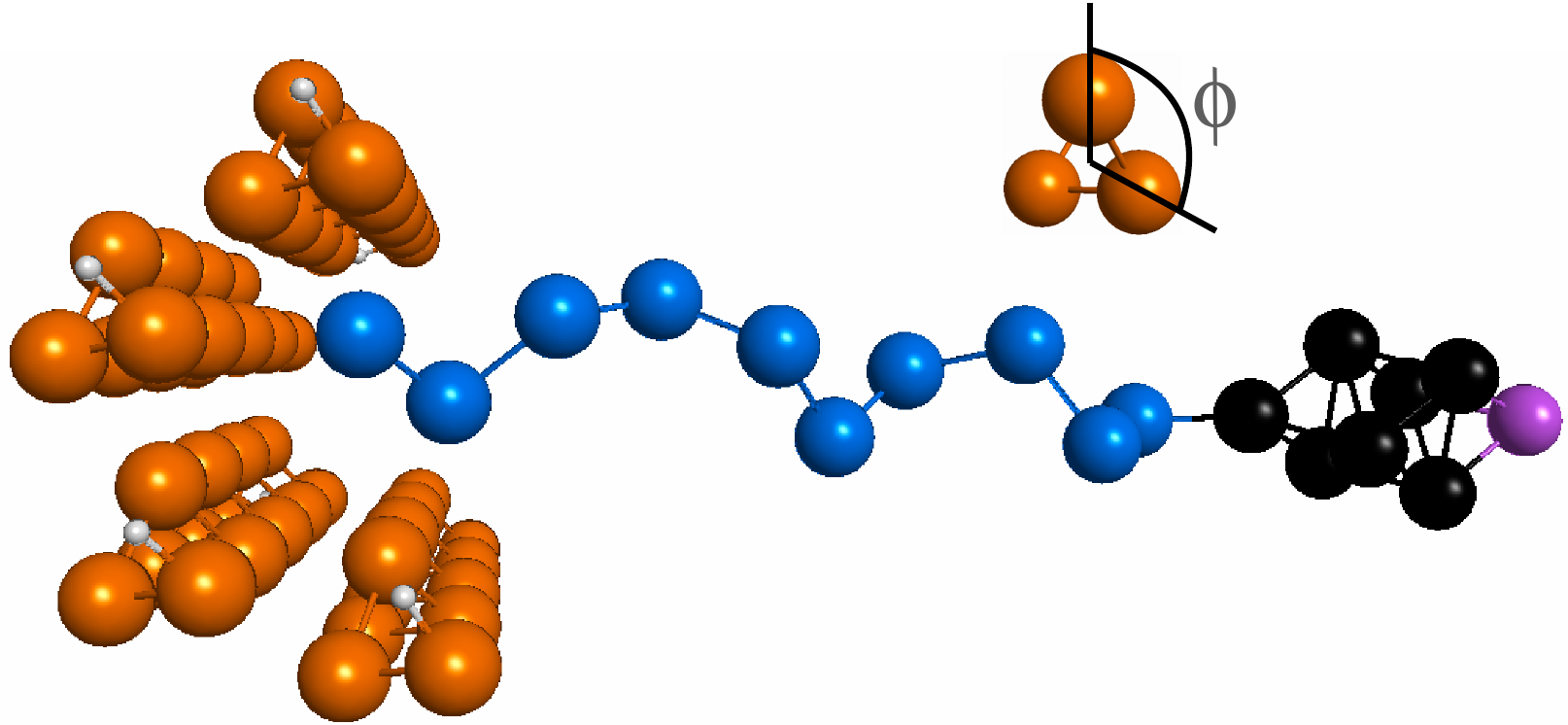}
\caption{(Color online).
Te atomic chain (blue) pulled from Te crystal surface (orange) by gold STM tip (black and mauve). Inset: $\phi$ is the twist angle per atom of a helix about its axis that is oriented perpendicularly to the page. Images prepared using Macmolplt software. \cite{MacMolPlt}
}
\label{end} 
\end{figure}

The gold STM tip is then backed away from the Te surface in small steps, optimizing the structure after each step while keeping the orange Te and mauve Au atoms frozen during each optimization.  The bonds between gold atoms, between gold and tellurium atoms, and intrachain bonds between tellurium atoms are all stronger than interchain bonds between tellurium atoms. Because of this it is possible for the blue atomic chain to be drawn out from the crystal surface by the STM tip without the Te chain or the STM tip or the bond beween them breaking. A resulting structure is shown in Fig.\ref{end}.

Te atomic helices can be characterized by their twist angles $\phi$ per atom about the axis of the helix,\cite{GK} defined in the inset of Fig.\ref{end}.
For the Te helices making up a Te crystal the twist angle is 120$^{\circ}$. Recently it has been predicted\cite{GK} that for single isolated long regular Te atomic helices the twist angle should be smaller, approximately 106$^{\circ}$. However, this prediction is yet to be tested experimentally. For the blue Te atomic chain shown in Fig.\ref{end} the helical twist angle varies between $\sim$102$^{\circ}$ and 110$^{\circ}$ except  near the Te crystal substrate.This is consistent with the above prediction,\cite{GK} that the twist angle for a free-standing Te helix should be substantially smaller than that for Te helices in Te crystals.

While the Te atomic chain is being stretched between the gold STM tip and the Te substrate it is under tension, which should inhibit its collapse into a compact disordered cluster such as those predicted in Ref.\onlinecite{GK}. 

 \section{Conclusion}
\label{C}

Although tellurium atomic chains encapsulated in a variety of hosts have been studied experimentally for more than 40 years, pristine free-standing Te atomic chains have still not been realized in the laboratory. Here it has been shown based on {\em ab initio} simulations that a pristine free-standing tellurium atomic helix can be drawn from a tellurium crystal surface with the help of a gold STM tip. While this helix is under tension it should not collapse into a tangled compact cluster. For the Te helix realised in this way the p-type semiconductor Te crystal substrate and gold STM tip provide suitable contacts for transport measurements on a pristine isolated helical Te atomic chain that have until now not been possible.

\begin{acknowledgments}
This research was supported by the Digital Research Alliance of Canada.
\end{acknowledgments}

{

\end{document}